\documentclass[conference]{IEEEtran}
\IEEEoverridecommandlockouts
\usepackage{cite}
\usepackage{amsmath,amssymb,amsfonts}
\usepackage{algorithmic}
\usepackage{listings}
\usepackage{color}
\usepackage[style=base]{caption}
\usepackage[pdftex]{graphicx,xcolor}
\usepackage{newtxtext}
\usepackage{graphicx}
\usepackage{subfigure}
\usepackage{epstopdf}
\usepackage{booktabs}
\usepackage{textcomp}
\def\BibTeX{{\rm B\kern-.05em{\sc i\kern-.025em b}\kern-.08em
    T\kern-.1667em\lower.7ex\hbox{E}\kern-.125emX}}
\begin{document}

\title{GraphEye: A Novel Solution for Detecting Vulnerable Functions Based on Graph Attention Network
}

\author{
\IEEEauthorblockN{1\textsuperscript{st} Li Zhou}
\IEEEauthorblockA{\textit{School of Information and Communication Engineering} \\
\textit{University of Electronic Science and Technology of China}\\
Chengdu, China \\
2018010801006@std.uestc.edu.cn}
\and
\IEEEauthorblockN{2\textsuperscript{nd} Minhuan Huang}
\IEEEauthorblockA{\textit{National Key Laboratory of Science and Technology} \\
\textit{on Information System Security}\\
Beijing, China \\
darbean@126.com}
\and
\IEEEauthorblockN{3\textsuperscript{rd} Yujun Li}
\IEEEauthorblockA{\textit{School of Computer Science and Engineering} \\
\textit{University of Electronic Science and Technology of China}\\
Chengdu, China \\
liyujun@uestc.edu.cn}
\and
\IEEEauthorblockN{4\textsuperscript{th} Yuanping Nie}
\IEEEauthorblockA{\textit{National Key Laboratory of Science and Technology} \\
\textit{on Information System Security}\\
Beijing, China \\
yuanpingnie@nudt.edu.cn}
\and 
\IEEEauthorblockN{5\textsuperscript{th} Jin Li}
\IEEEauthorblockA{\textit{National Key Laboratory of Science and Technology} \\
\textit{on Information System Security}\\
Chengdu, China \\
201922081124@std.uestc.edu.cn}
\and
\IEEEauthorblockN{6\textsuperscript{th} Yiwei Liu}
\IEEEauthorblockA{\textit{School of Computer Science and Engineering} \\
\textit{University of Electronic Science and Technology of China}\\
Chengdu, China \\
2017060901015@std.uestc.edu.cn}
}


\maketitle

\begin{abstract}
With the continuous extension of the Industrial Internet, cyber incidents caused by software vulnerabilities have been increasing in recent years. However, software vulnerabilities detection is still heavily relying on code review done by experts, and how to automatedly detect software vulnerabilities is an open problem so far. In this paper, we propose a novel solution named GraphEye to identify whether a function of C/C++ code has vulnerabilities, which can greatly alleviate the burden of code auditors. GraphEye is originated from the observation that the code property graph of a non-vulnerable function naturally differs from the code property graph of a vulnerable function with the same functionality. Hence, detecting vulnerable functions is attributed to the graph classification problem.GraphEye is comprised of VecCPG and GcGAT. VecCPG is a vectorization for the code property graph, which is proposed to characterize the key syntax and semantic features of the corresponding source code. GcGAT is a deep learning model based on the graph attention graph, which is proposed to solve the graph classification problem according to VecCPG. Finally, GraphEye is verified by the SARD Stack-based Buffer Overflow, Divide-Zero, Null Pointer Deference, Buffer Error, and Resource Error datasets, the corresponding F1 scores are 95.6\%, 95.6\%,96.1\%,92.6\%, and 96.1\% respectively, which validate the effectiveness of the proposed solution.
\end{abstract}


\section{Introduction}
Software vulnerabilities refer to software design or implementation defects, which may be exploited by malicious users to achieve information leakage, resource utilization, and facility destruction. In recent years, with the rapid development of the Industrial Internet and continuously emerging applications, functionalities of different software have become more and more complex and larger in scale. In addition to the complexity of source code quality management, zero-day and n-day software vulnerabilities have shown an upward trend. Thus, cyber incidents caused by this kind of vulnerability have also increased. Hence, software vulnerability detection based on source code, as an old research topic, has once again received significant attention recently\cite{Russell@2018,Young-Su@2018,lizhen,LSTM,Wang@2021,sysevr}.

The mainstream method of software vulnerability detection based on source code is to convert firstly the source code into an abstract representation and then analyze the abstract representation to check whether it matches a certain predefined vulnerability detection rule, to determine finally whether the source code contains the corresponding vulnerabilities\cite{lizhen}. According to specific analysis techniques, the vulnerability detection methods can be divided into three categories: code similarity-based vulnerability detection, pattern-based vulnerability detection, and machine learning-based vulnerability detection.

Code similarity-based vulnerability detection originates that similar codes are likely to contain the same vulnerabilities. Code segments are abstractly represented based on their characteristics, and then judge the similarity between the code to be detected and the code containing a known vulnerability according to their corresponding representations, and determine finally whether the detected code contains the corresponding vulnerability. Based on the above idea, ReDeBug can quickly find unpatched code clones in OS-distribution scale code bases\cite{ReDeBug}, VulPecker can automatically detect whether a piece of software source code contains a given vulnerability or not\cite{VulPecker}, VUDDY can fully detect security vulnerabilities in large software programs efficiently and accurately by leveraging function-level granularity and a length-filtering technique\cite{VUDDY}, and so on. The principle of code similarity-based vulnerability detection is clear and easy to understand, but this approach is limited to detecting vulnerabilities incurred by code cloning or approximate code cloning, and the false negatives for non-code-cloning vulnerabilities are high.

The core of pattern-based vulnerability detection is plenty of rules formulated by lots of analysts based on domain knowledge, historical vulnerability data, or vulnerable codes directly. Each rule can capture the essential characteristics of a kind of vulnerability from the abstract representation of the corresponding source code, and hence can be used to identify whether the detected code contains the specific vulnerabilities. This method is widely used in current tools for automatic code analysis, including open-source software such as Flawfinder, RATS, and ITS4, and commercial software such as Checkmarx, Fortify, and Coverity. Recently, Yamaguchi et al. advance this approach by a novel representation of source code that named the code property graph which merges concepts of abstract syntax trees, control flow graphs, and program dependence graphs. Based on this representation, common vulnerabilities can be modeled as graph traversals which can identify buffer overflows, integer overflows, format string vulnerabilities, or memory disclosures\cite{modeling}. Pattern-based vulnerability detection methods can accurately locate the vulnerability but are rather laborious. Furthermore, this kind of method may lead to both a high false-positive rate and a high false-negative rate due to imperfect rules and are entirely incapable of unknown vulnerabilities.

Machine learning-based vulnerability detection is proposed to reduce the reliance on domain experts, which can be subdivided into traditional machine learning-based vulnerability detection and deep learning-based vulnerability detection according to whether domain experts are required to define features. Traditional machine learning-based vulnerability detection relies on domain experts to manually define features, and use machine learning models, such as KNN, SVM, C4.5, and so on, to automatically classify vulnerable code and non-vulnerable code. For example, to detect subtle taint-style vulnerabilities from C source code, Yamaguchi et al. introduce unsupervised machine learning to construct patterns that are usually identiﬁed by manual analysis\cite{automatic}. Better than traditional machine learning-based vulnerability detection, deep learning-based vulnerability detection can automatically generate vulnerability patterns, which alleviates the requirements to manually define features further. Zhen Li et al. propose the first systematic framework for using deep learning to detect vulnerabilities in C/C++ programs with source code just recently\cite{sysevr}.

Deep learning has been successful in the image and natural language process and is very promising in vulnerability detection. However, most of the recent works focus on how to apply traditional vectorization methods in natural language processing, such as word2vec, glove, and so on, to the program source code\cite{LSTM,sysevr,smartContract}. However, the program source code differs from the image and natural language in nature. More efforts for vectorization of program source code are needed to improve deep learning vulnerability detection. Furthermore, research is also needed in terms of detection accuracy, vulnerability location, large-scale labeled datasets, model interpretation, and so on.

\textbf{Our contributions.} In this paper, we firstly propose that detecting vulnerable functions of c/c++ code is attributed to the graph classification problem. Then, a novel solution named GraphEye for this problem is proposed. GraphEye is comprised of a vectorization for the code property graph and a deep learning model based on the graph attention network. The focus of this paper is centered on answering the following question: \textit{How can we detect vulnerable functions based on graph neural network model, given the fact that the code property graphs of these functions have fully captured enough syntax and semantic information to identify the potential vulnerabilities?}

The remainder of the paper is organized as follows: Section 2 briefly introduces the code property graph. Section 3 describes the solution framework in detail. Section 4 analyzes experiment results in depth. Section 5 concludes our work and discusses the future directions.

\section{Code Property Graph Overview}

A graph can be formally defined as $G=(V, E)$ in math theory, where $V$ is a set of nodes, and $E \subseteq (V \times V)$ is a set of edges. However, this highly abstract definition ignores the fact that there may be significant differences both between entities and the relationships between entities in the real world. Hence, the concept of property graph comes into being, which is an extension of traditional graph definition by characterizing nodes and edges' properties. A property graph is a directed labeled multigraph with the special characteristic that each node or edge could maintain a set (possibly empty) of property-value pairs\cite{graphModel}. The definition of property graph can be described as following:

\textbf{Definition 1:} A property graph is a five-tuple $G=(V,E,\lambda,\Gamma,\mu)$ , where $V$ is a set of nodes, $E \subseteq (V \times V)$ is a set of directed edges from source node to destination, $\lambda$ is a labeling function for nodes, $\Gamma$ is a type function for edges and $\mu$ is a property function for both nodes and edges.

\begin{figure}[tb]
\begin{center}
\includegraphics[width=8cm]{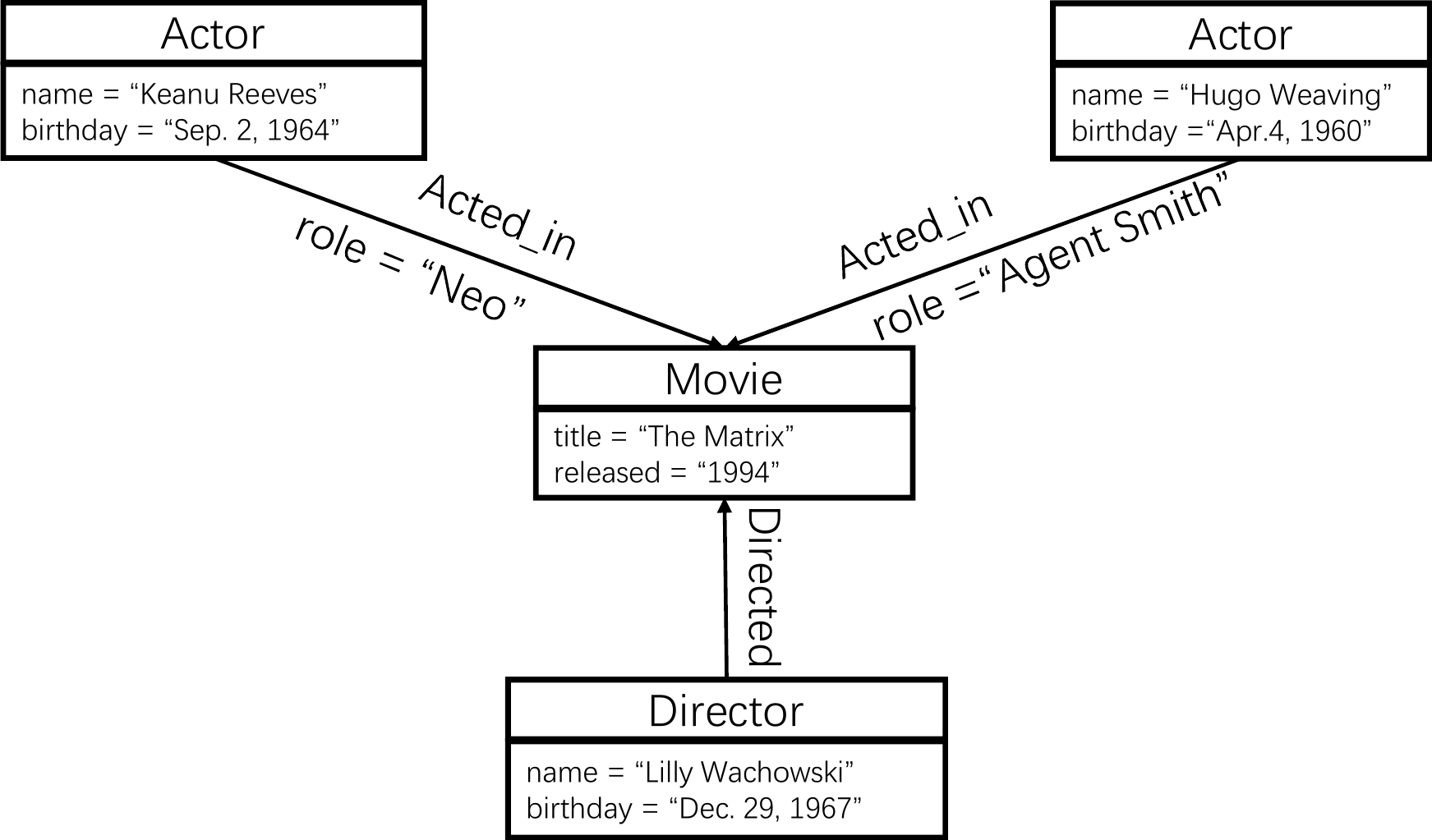}
\end{center}
\caption{a simple movie property graph}
\label{fig:1}
\end{figure}

In a property graph, each node has at most one label, and each edge has at most one type, respectively identifying the classes of nodes and edges. Any node or edge in the property graph can have zero or more attributes to identify the characteristics of the node or edge. Figure~\ref{fig:1} illustrates a typical property graph about a movie, and the charactersistics of $v_1$ and $e_1$ can be described by functions as follows:

\begin{align*}
& \lambda(v_1) = \{Actor\} \\
& \mu(v_1, name) = "Keanu\ Reeves" \\
& \mu(v_1, birthday) = "Sep.\ 2,\ 1964" \\
& \Gamma(e_1) = \{Acted\_\ in\ \} \\
& \mu(e_1, role) = "Neo" \\
\end{align*}

With the help of the property graph, Yamaguchi et al. firstly merge the concept of the abstract syntax tree, control flow, and program flow chart to form a novel representation of program source code that named the code property graph in their publication\cite{modeling}. The code property graph can be formally defined as follows:
\begin{align*}
G_{CPG} &=(V_{CPG},E_{CPG},\lambda_{CPG},\Gamma_{CPG},\mu_{CPG} ) \\
&= G_{AST} \cup G_{CFG} \cup C_{PDG}
\end{align*}
where $G_{AST}$, $G_{CFG}$ and $G_{PDG}$ are the representation with property graph for the traditional abstract syntax tree, the control flow graph, and the program dependence graph of a program source code respectively.

This combination is prior to a single representation alone to characterize a vulnerability type in the vast majority of cases. As mentioned in \cite{patten-based} that the code property graph is not limited to the abstract syntax tree, the control flow graph, and the program dependence graph, more additional representations can be overlaid to extend the capability of the code property graph. We also note that more traditional representations, such as data dependence graph and control dependence graph, have been merged into the code property graph in Joern\cite{joern} which is an open-source tool to generate the code property graph.

\lstset{
language=C++,
basicstyle=\small\ttfamily,
numbers=left,
numbersep=5pt,
xleftmargin=20pt,
frame=tb,
breaklines=true,
extendedchars=true,
framexleftmargin=20pt
}

\renewcommand*\thelstnumber{\arabic{lstnumber}:}

\captionsetup[lstlisting]{labelfont=bf,singlelinecheck=off,labelsep=space}

\begin{lstlisting}[caption={A bad function with divide-zero error}, label={code:3}]
static void bad(float Data)
{
float data = Data;
{
/* POTENTIAL FLAW: Possibly divide by zero */
int result = (int)(100.0/data);
printIntLine(result);
}
}
\end{lstlisting}

\begin{figure}[tb]
\begin{center}
\includegraphics[width=8cm]{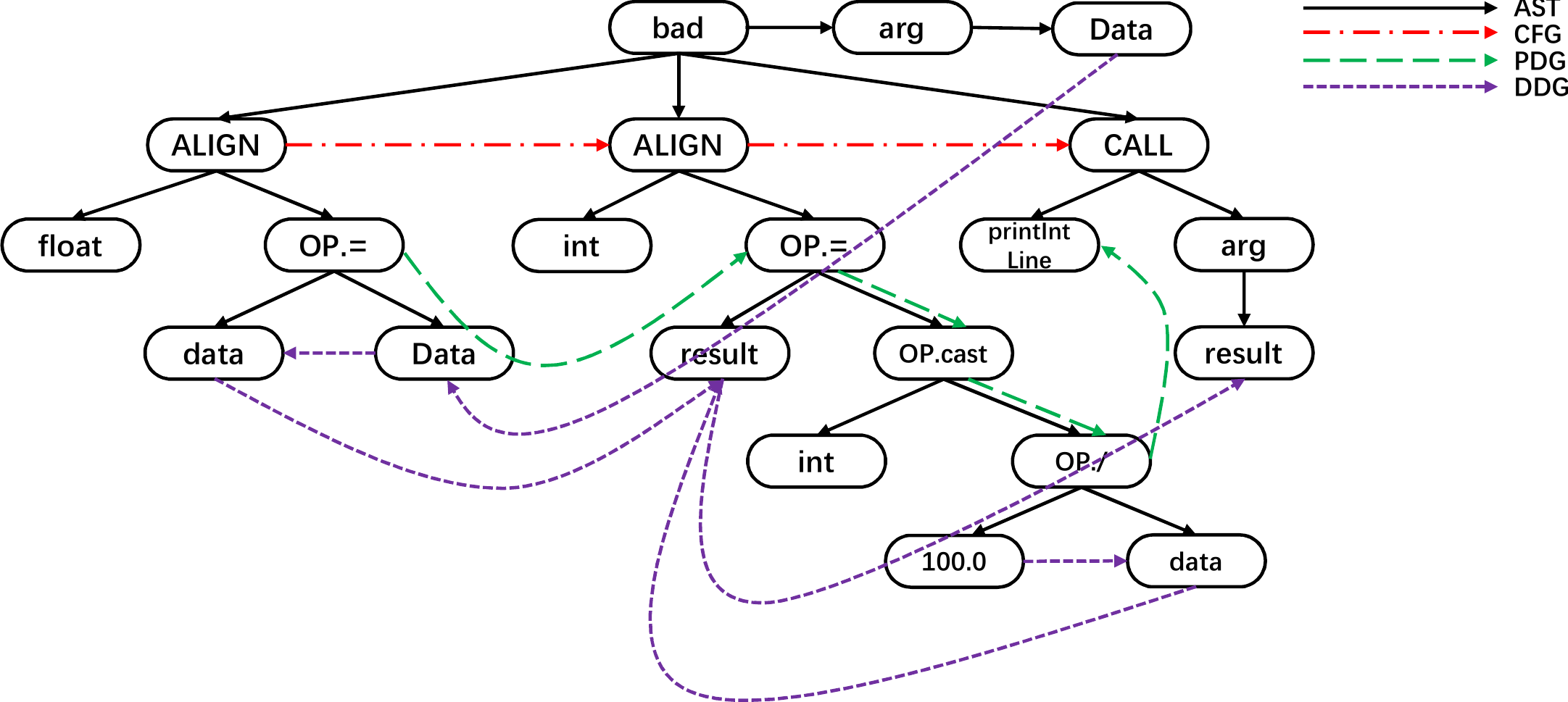}
\end{center}
\caption{The code property graph of bad() function}
\label{fig:7}
\end{figure}

For instance, Fig. ~\ref{fig:7} illustrates the code property graph of a simple function depicted in Listing~\ref{code:3} from Juliet\cite{juliet}. In Fig.~\ref{fig:7}, the edges of the abstract syntax tree, control flow graph, program dependence graph, and data dependence graph are indicated by the black solid lines, the red dashed lines, the purple dotted lines, and the green dashed-dotted lines respectively. All the syntax structs such as variable, data type, operate, statement, the function call, and so on are included in the subgraph consisting of black solid edges and the corresponding nodes. All the control flow information, i.e., the execution order of statements, described by the subgraph consisting of the red dashed edges and the corresponding nodes. All the control dependencies are depicted in the subgraph consisting of the purple dotted edges and the corresponding nodes. All the data dependencies are illustrated by the subgraph comprising of the green dashed-dotted edges and the corresponding nodes.
\section{Solution Framework}
\subsection{Motivation and Overview}
Our motivation comes from both the capture of vulnerability characteristics by the code property graph and the development of graph neural network technology. We first observe that the code property graph of the non-vulnerable source code differs naturally from the code property graph of the vulnerable source code with the same functionality. Then, we also notice that graph neural networks have been used for graph classification\cite{imbalance_learning}.

\begin{figure}[tb]
\begin{center}
\includegraphics[width=8cm]{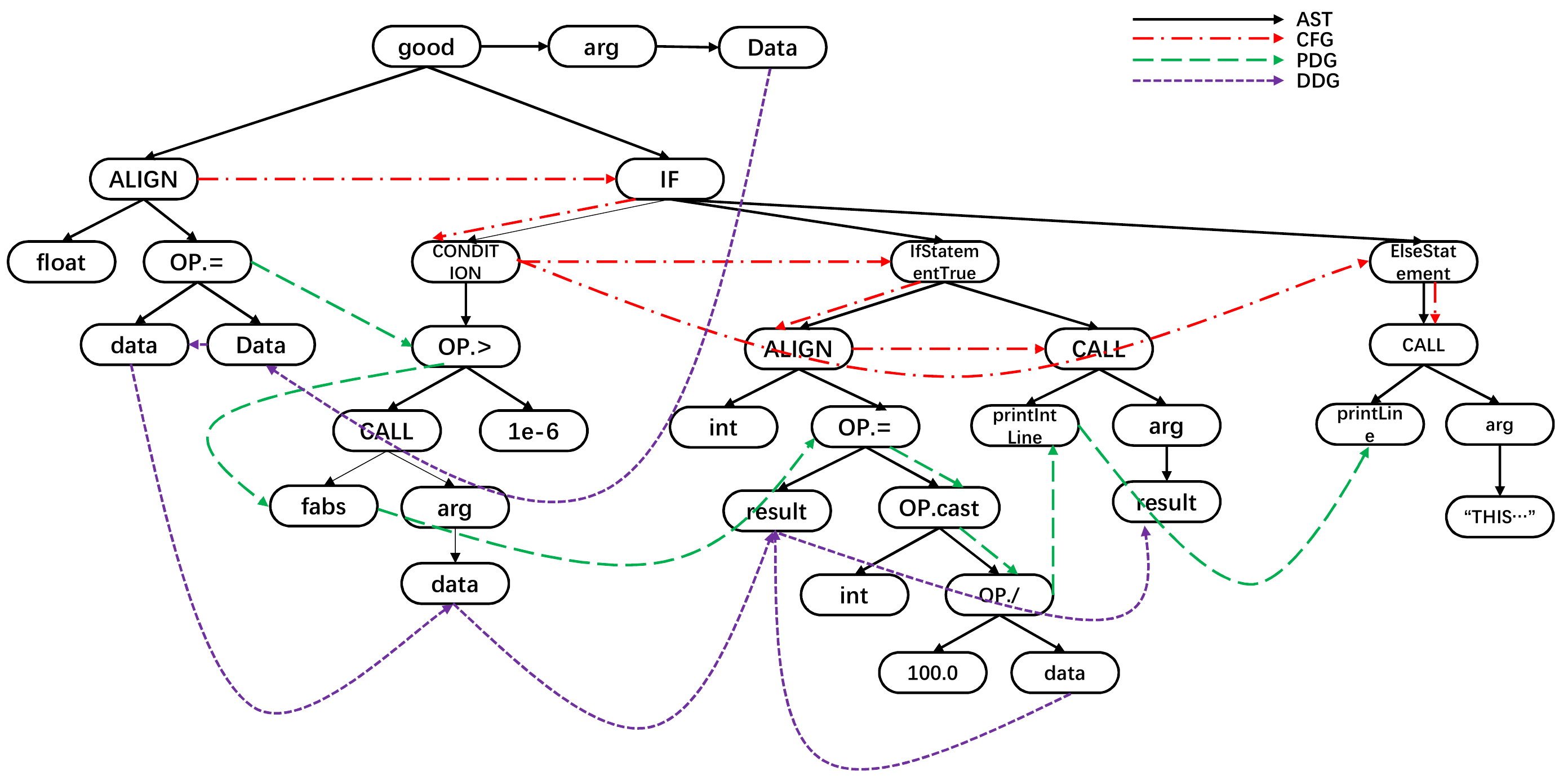}
\end{center}
\caption{The code property graph of good() function}
\label{fig:6}
\end{figure}

The differences in the code property graph of the non-vulnerable source code and the vulnerable can be illustrated by the following example. The fixed code of bad() function in Listing~\ref{code:3} is depicted in Listing \ref{code:2}, and its code property graph is shown in Fig.~\ref{fig:6}. There is a lack of a subgraph to judge whether data is equal to zero in Fig. \ref{fig:7}, and the judgment is essential to lead to the error of divided by zero. It must be pointed out that the difference is not limited to the above vulnerability type, and it does exist in all vulnerability types as long as that the code property graph is enough overlayed.

\begin{lstlisting}[caption={A good funciotn without divide-zero error}, label={code:2}]
static void good(float Data)
{
float data = Data;
if(fabs(data) > 0.000001)
{
int result = (int)(100.0/data);
printIntLine(result);
}
else
{
printLine("This would result in a divide by zero");
}
}
\end{lstlisting}

\begin{figure*}[tb]
\begin{center}
\includegraphics[width=16cm]{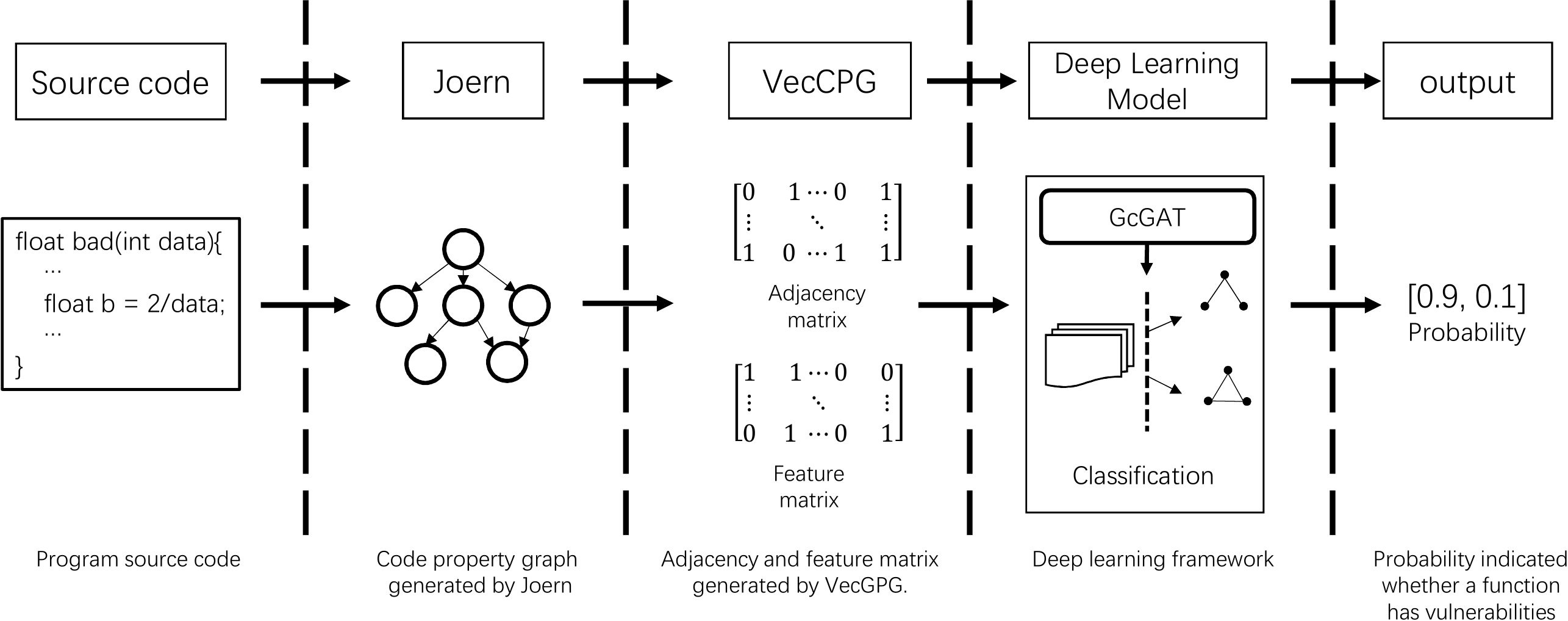}
\end{center}
\caption{The solution framework}
\label{fig:10}
\end{figure*}

Hence, detecting vulnerable functions of c/c++ code is modeled as the graph classification problem, and the solution framework is illustrated as Fig.\ref{fig:10}.The solution framework can be divided into three components. The first component is the generation of the code property graph for the program source code, which is the basis of our works and can be done by Joern. The second component is the vectorization of the property graph, and which is the foundation for the application of graph neural network model and can be done by our novel schema called VecCPG (\textbf{\emph{\underline{Vec}}}torization for the \textbf{\emph{\underline{C}}}ode \textbf{\emph{\underline{P}}}roperty \textbf{\emph{\underline{G}}}raph of a program source code). The third component is the deep learning model, and we propose a novel model, named GcGAT (\textbf{\emph{\underline{G}}}raph \textbf{\emph{\underline{C}}}lassification based on \textbf{\emph{\underline{G}}}raph \textbf{\emph{\underline{A}}}ttention Ne\textbf{\emph{\underline{T}}}works), to detect vulnerable functions. The combination of VecCPG and GcGAT is called GraphEye, which is the core of the solution framework.

\subsection{Vectorization}
Although the extension of the code property graph for a program source code has captured most vulnerability types so far, there is still a gap that must be filled before the model of graph neural network can be used to detect vulnerable functions. That is to say, how to vectorize the code property graph? VecCPG is proposed to fill the gap.VecCPG is comprised of the feature matrix and the adjacent matrix. The feature matrix represents nodes’ information, which captures the syntax characteristics of a program source code. For a given code property graph $G_{CPG} =(V_{CPG},E_{CPG},\lambda_{CPG},\Gamma_{CPG},\mu_{CPG} )$, the feature matrix of $G_{CPG}$ is defined as follows:
\begin{center}
\begin{equation*}
X = R^{|V_{CPG}|\times |F|}
\end{equation*}
\end{center}
where $|V_{CPG}|$ is the cardinalities of the node-set $V_{CPG}$, and $|F|$ is the dimension of the selected properties of nodes. $F$ represents the features related to vulnerabilities, which is consisted of five different components illustrated in Table \ref{table:1}.

\begin{table}[htb]
\caption{The structure of VecCPG.}
\label{table:1}
\begin{center}
\begin{tabular}{ccccc}
\toprule
label&operator&function&literal&type\\
\hline
$13+2$&$25+2$&$39+2$&$32$&$16+2$\\
\bottomrule
\end{tabular}
\end{center}
\end{table}

The structure of VecCPG characterizes the syntax details of the node label, operator, API function call, constant, and variable type.

\textbf{Label:} This label indicates which class a node belongs to, which is similar to the meaning of a label in a traditional property graph. There are 13 different classes are considered in this paper, and all the labels and their corresponding implications are listed in table \ref{table:2}. All these labels are encoded in a one-hot way, one additional bit for the unknown node, another additional bit for reservation.

\begin{table}[htb]
\caption{Labels and implications.}
\label{table:2}
\begin{center}
\begin{tabular}{lp{4cm}}
\toprule
\textbf{Label} & \textbf{implications} \\
\hline
INDENTIFIER & variables\\
\hline
LITERAL & constants, such as strings, integers\\
\hline
LOCAL & variables in the function body that has been declared\\
\hline
BLOCK & separator\\
\hline
METHOD\_RETURN & the return of a method\\
\hline
METHOD & a method definition \\
\hline
CONTROL\_STRUCTURE & the control statement structure, such as \textit{if}, \textit{while}\\
\hline
FIELD\_IDENTIFIER & a reference to a namespace, usually is \textit{::}\\
\hline
UNKNOWN & unknown types\\
\hline
RETURN & the return of a function\\
\hline
PARAM & the parameters of a function\\
\hline
JUMP\_TARGET & the label used by goto\\
\hline
CALL & a call to a function or operator\\
\bottomrule
\end{tabular}
\end{center}
\end{table}

\textbf{Operator:} 25 types of operators including arithmetic operators, rational operators, logical operators, bitwise operators, pointer operators, and so on are considered in these components. The detailed operators and the corresponding meaning are described in Table \ref{table:3}. Operators are also encoded in a one-hot way, one additional bit for the unknown node, another additional bit for reservation.

\begin{table}[htb]
\caption{Operators and implications.}
\label{table:3}
\begin{center}
\begin{tabular}{lr}
\toprule
\textbf{operator} & \textbf{meaning}\\
\hline
= & assignment \\
\hline
\textit{[]} & indirectIndexAccess \\
\hline
sizeof() & sizeOf \\
\hline
* & multiplication \\
\hline
() & cast \\
\hline
- & subtraction \\
\hline
. & fieldAccess\\
\hline
< & lessThan \\
\hline
++ & postIncrement \\
\hline
\& & addressOf \\
\hline
+ & addition \\
\hline
== & equals \\
\hline
** & indirection \\
\hline
- & minus \\
\hline
!= & notEquals \\
\hline
>= & greaterEqualsThan \\
\hline
-> & indirectFieldAccess \\
\hline
| & logicalOr \\
\hline
/ & division \\
\hline
\& & logicalAnd \\
\hline
delete() & delete \\
\hline
\&\& & and \\
\hline
> & greaterThan \\
\hline
\% & modulo \\
\hline
new & new\\
\bottomrule
\end{tabular}
\end{center}
\end{table}

\textbf{Function:} Considering some API function calls, for example, memcpy(), may lead to vulnerabilities, hence API function names are encoded into VcCPG in a one-hot way. The number of different API functions varies in different datasets. Thus, the simplest way to encode all API functions for a certain language. However, Limited to the dataset used in this paper, 39-bit encodes are enough to represent all API functions, one additional bit for unknown, another additional bit for reservation. Surely, the encoding is easily extended to meet requirements in real application scenarios.

\textbf{Literal:} Some constants are also factors for the vulnerabilities, such as the divide-by-zero vulnerability. Due to the fact that the dataset in this paper is limited to 32 bits, all integers are encoded into 32 bits same to the underlying storage method of the system, and the floating-point constants are encoded according to IEEE 754 standard\cite{IEEE754}. Of course, this encoding is easily extended to 64-bit systems.

\textbf{Type:} For C/C++ languages,10 basic variable types and 6 complex variable types are considered. These basic variable types are char, int, short, float, double, long, string, void, struct, and union. Those complex variable types are signed, unsigned, *, array, map, and vector. A basic variable type and a complex variable type can be combined together, such as “char *”. Basic variable types and complex variable types are independently encoded in a one-hot way with an additional one-bit reservation.

The adjacency matrix is composed of AST, CFG, and DDG edges in the code property graph of a program source code, reflecting semantic information such as dependence and control between nodes. The adjacency matrix is defined as $A= R^{(|V_{CPG}|\times|V_{CPG}|)}$ where $|V_{CPG}|$ is the cardinalities of the node-set $V_{CPG}$. Compared to the feature matrix, the adjacent matrix is easier to construct. As long as there is at least one edge between two nodes, regardless of the type and number of edges, the element in the corresponding adjacency matrix is set to 1; otherwise, it is set to 0.

\subsection{Deep Learning Model}
\begin{figure*}[tb]
\begin{center}
\includegraphics[width=16cm]{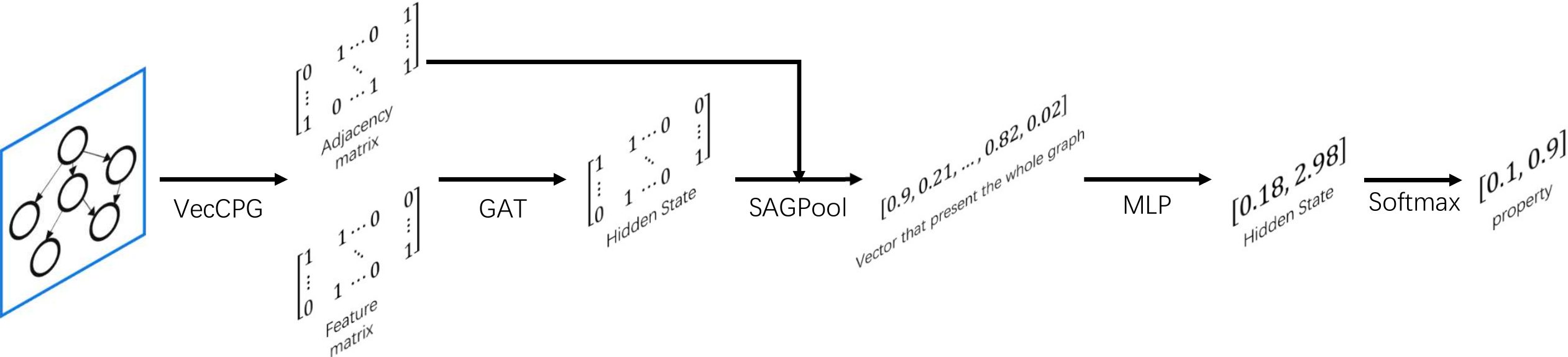}
\end{center}
\caption{GcGAT}
\label{fig:11}
\end{figure*}

A novel deep learning model, named GcGAT is proposed to detect vulnerable functions of the program source code. This model illustrated in Fig.\ref{fig:11} includes GAT, SAGpool, MLP, and softmax.

In traditional graph attention network application scenarios, the input is a feature matrix and an adjacency matrix. After the feature extraction of the multi-head attention model, the feature vectors of all nodes are output to label different nodes. However, what we need is to classify graphs rather than labeling nodes in a graph. If GAT's output feature matrix is directly expanded into a vector and then input into MLP for classification, it may lead to the problem of excessive MLP parameters and the consequent high-dimensional curse. Inspired by the convolutional layer and pooling layer of CNN which is introduced to solve the classification problem by converting the output feature matrix of the graph into a vector. An improved SAGPool after GAT is introduced as our graph pooling.

Traditional SAGPool is an implementation of hierarchical pooling. It adaptively learns the importance of nodes from the graph through graph convolution and discards non-important nodes based on the TopK mechanism. Instead, we use GCN to directly reduce the dimensionality of the feature matrix output by GAT and then convert it into a vector to represent the whole graph. After obtaining the feature vector of the graph, we input it into MLP with fewer layers and parameters for classification. The number of nodes in the input layer of MLP is the dimension of the feature vector of the graph, and the number of nodes in the output layer of MLP is 2, which indicates that the results are divided into two categories. Finally, a common function named softmax is introduced for normalization to meet requirements of probability for MLP’s output. The definition of softmax is as follows:

\begin{equation*}
\begin{aligned}
S_i = \frac{e^{z_i}}{\sum^{K}_{j=1}e^{z_j}}\\
\end{aligned}
\end{equation*}

where $z_i$ is the $i_{th}$ element and $S_i$ is the corresponding output of softmax.

\section{Experiments}
\subsection{Data Preprocessing}
SARD\cite{SARD} is a dataset of different types of vulnerabilities, which has been widely used for researchers to evaluate their methods. In terms of model training and testing, we have selected three classic sub-datasets, namely, CWE 121 Stack-based Buffer Overflow, CWE 369 Divide-Zero, CWE 476 NULL Pointer Deference. In comparison with other models, we have selected two more widely used vulnerability types: CWE 199 Buffer Error and CWE-399 Resource Management Error.

Remember that this paper discusses whether there are vulnerabilities in functions. Hence, for training and testing the proposed deep learning model, the first thing that needs to do is to label the functions in the dataset classified by whether a function has vulnerabilities. At first, the concept of root bad function and root good function is introduced as follows:

\textbf{Root bad function:} The function that its vulnerabilities are caused by either the statements themselves or API function calls is a root bad function. That is to say, the vulnerabilities in this function are not caused by user-defined function calls.

\textbf{Root good function:} The function that has no vulnerabilities or its vulnerabilities are caused by user-defined function calls is a root good function. That is to say, there are no vulnerabilities under the exception of user-defined function calls.

\begin{table}[htb]
\caption{The distribution of labeled functions}
\label{table:5}
\begin{center}
\begin{tabular}{|l|l|l|l|l|}
\hline
\textbf{CWE} & \textbf{Name} & \textbf{good} & \textbf{bad} & \textbf{Total}\\
\hline
121 & Stack-based Buffer Overflow & 529 & 3643 & 4172 \\
\hline
369 & Divide-Zero & 845 & 593 & 1368 \\
\hline
476 & NULL Pointer Dereference & 410 & 274 & 684\\
\hline
119 & Buffer Error & 844 & 4682 & 5330 \\
\hline
399 & Resource Error & 1179 & 802 & 1915 \\
\hline
\end{tabular}
\end{center}
\end{table}

Then, based on the above definitions, all the functions without user-defined function calls are picked up and labeled “good” or “bad” respectively in the dataset, as illustrated in Table \ref{table:5}.

\subsection{Training Process}
Since the feature matrix of a code property graph is relatively sparse, the gradient update needs more epochs to obtain better parameters. A module named ray.tune is applied to perform a grid search on hyperparameters and pick up the best hyperparameters. Then based on the best hyperparameters, training by some epochs for the fixed dataset. Generally, the model is seriously underfitting from the 1st to 5th epochs. By the 7th epoch, the model performance has been greatly improved. After 10 epochs, the model performance has stabilized, and the training can be considered to be over. The hyperparameters finally used in GcGAT are as follows: learning rate equals 8.6e-4, the number of epochs is 15; the dropout is 0.3; the dimension of the hidden layer is 64 and the dimension of the pooling layer is 32.

Due to the imbalance problem of the dataset under the two classifications\cite{imbalance_learning}, a penalty factor strategy is adopted in GcGAT. For frequently occurring classes, the penalty is reduced by multiplying a number less than 1. For losses in small samples, the penalty is increased by multiplying a number greater than 1. In our experiments, the penalty factor is optimized to 0.6 for frequently occurring classes, and the penalty factor is optimized to 1.7 for small samples by hyperparameters searching.

For CWE 121 Stack-based Buffer Overflow, the large-sample down-sampling technique is adopted to balance the positive samples and negative samples. The large-sample down-sampling technique is not adopted for CWE 369 Divide-Zero and CWE 476 NULL Pointer due to their balanced samples. CWE 119 Buffer Error and CWE 399 Recourse Error are used for comparisons with other methods, the large-sample down-sampling technique is not adopted to ensure fairness. For all types of vulnerabilities, the dataset is randomly divided into the training set and test set according to the ratio of 8:2.

We implement our framework in Python using Pytorch. The computer running experiments has two NVIDIA RTX TITAN GPUs and an Intel Xeon E5-2678 v3 CPU running at 3.30GHz.When training the neural networks to find vulnerabilities, the framework only consumes 2.46G memory with 1.16\% average load on CPU, and 1.4G GPU memory with 29\% average load on GPU. And when we leverage our model for inference, the speed reaches 185.69 functions per second on average, which indicates that our model can detect vulnerabilities rapidly under the condition of low resource usage.

\subsection{Experiment Results}
FPR (False Positive Rate), FNR (False Negative Rate), TPR (True Positive Rate), P (Precision) and F1 are common metrics for evaluting the effectiveness of a deep learning model. In terms of software vulnerabilities detection,their definitions can be described as follows:
\begin{align*}
&FPR = \frac{FP}{FP + TN} \\
&FNR = \frac{FN}{TP + FN} \\
&TPR = \frac{TP}{TP + FN} \\
&P = \frac{TP}{TP + FP} \\
&F1 = \frac{2 \times P \times TPR}{P + TPR}
\end{align*}
where $FP$ denotes the number of samples that are non-vulnerable but detected as vulnerable, $TN$ denotes the number of samples that are non-vulnerable and detected as non-vulnerable, $FN$ denotes the number of samples that are vulnerable but detected as non-vulnerable, and $TP$ denotes the number of samples that are vulnerable and detected as vulnerable. It is clear that the lower FRR and FNR, the higher TPR, P and F1 implicates the better effectiveness of the model.

\begin{table}[tb]
\caption{The comparison with others for CWE 119}
\label{table:7}
\begin{center}
\begin{tabular}{|l|l|l|l|l|l|}
\hline
\textbf{ } & \textbf{FPR} & \textbf{FNR} & \textbf{TPR} & \textbf{P} & \textbf{F1}\\
\hline
Flawfinder & 56.6\% & 44.8\% & 55.2\% & 39.9\% & 46.3\% \\
\hline
RATS & 68.7\% & 31.3\% & 68.7\% & 40.5\% & 51.0\% \\
\hline
Paper\cite{LSTM} & 14.3\% & 14.6\% & 85.4\% & 80.4\% & 82.8\% \\
\hline
GraphEye & 9.4\% & 12.2\% & 87.8\% & 98.0\% & 92.6\% \\
\hline
\end{tabular}
\end{center}
\end{table}

\begin{table}[tb]
\caption{The comparison with others for CWE 399}
\label{table:8}
\begin{center}
\begin{tabular}{|l|l|l|l|l|l|}
\hline
\textbf{ } & \textbf{FPR} & \textbf{FNR} & \textbf{TPR} & \textbf{P} & \textbf{F1}\\
\hline
Flawfinder & 40.7\% & 58.4\% & 41.6\% & 34.1\% & 37.4\% \\
\hline
RATS & 33.9\% & 63.8\% & 36.2\% & 35.0\% & 35.6\% \\
\hline
Paper\cite{LSTM} & 14.1\% & 18.5\% & 81.5\% & 73.7\% & 77.4\% \\
\hline
GraphEye & 0.4\% & 7.0\% & 93.0\% & 99.3\% & 96.1\% \\
\hline
\end{tabular}
\end{center}
\end{table}

For CWE 119 Buffer Error and CWE 399 Recourse Error, the experiment results compared with Flawfinder, RATS, and Paper\cite{LSTM} are shown in Table \ref{table:7} and Table \ref{table:8}. Results of Flawfinder, RATS and Paper\cite{LSTM} come from \cite{LSTM}. Both FlawFinder and RATS are vulnerability detection tools based on pattern recognition, while Paper\cite{LSTM} adopts deep learning models to detect vulnerabilities. It is obvious from Table \ref{table:7} and Table \ref{table:8} that the effectiveness of Flawfinder and RATS is poor. This is because the predefined rules for vulnerabilities are usually simple and are difficult to handle complex and flexible implementation. The effectiveness of Paper\cite{LSTM} has been greatly improved, and far better than Flawfinder and RATS in all $FPR$,$FNR$,$TPR$,$P$, and $F1$ metrics. Unlike natural language processing adopted in Paper\cite{LSTM}, GraphEye is rooted in graph attention network and can capture more syntax structure and semantic information characteristics from the program source code. Hence, the overall effectiveness of GraphEye is much better than Paper\cite{LSTM}.

\begin{table}[tb]
\caption{The experiment results for CWE121, CWE369 and CWE476}
\label{table:6}
\begin{center}
\begin{tabular}{|l|l|l|l|l|l|}
\hline
\textbf{ } & \textbf{FPR} & \textbf{FNR} & \textbf{TPR} & \textbf{P} & \textbf{F1}\\
\hline
CWE 121 & 2.9\% & 6.7\% & 93.3\% & 97.9\% & 95.6\% \\
\hline
CWE 369 & 4.4\% & 2.7\% & 97.3\% & 93.9\% & 95.6\% \\
\hline
CWE 476 & 1.3\% & 5.8\% & 94.2\% & 98.0\% & 96.1\% \\
\hline
\end{tabular}
\end{center}
\end{table}
The experiment results for CWE 121 Stack-based Buffer Overflow, CWE 369 Divide-Zero, and CWE 476 NULL Pointer Deference are shown in Table \ref{table:6}. F1 of these three types of vulnerabilities are all higher than $95\%$.For CWE 121 Stack-based Buffer Overflow, the difference between a root good function and a corresponding root bad function mainly lies in the value of the variable or constant, rather than the logical structure of the program source code. While GAT is sensitive to structure but is a little obtuse to constants. Hence, GraphEye adopts the penalty for bad function samples in the training process, which leads to the high false alarm rate. For CWE 369 Divide-Zero and CWE 476 NULL Pointer Deference, the results are similar. This is because that these two types of vulnerabilities mainly depend on whether there are branch statements for different values of variables. Some useless structures such as if(0) can Interference GraphEye, so the false-negative rate is higher.

\section{Conclusion and Future Work}
This paper firstly proposes that detecting vulnerable functions can be attributed to the graph classification problem. Then, a novel solution named GraphEye for this problem is proposed. GraphEye is comprised of VecCPG and GcGAT. VecCPG is a vectorization for the code property graph, which reflects the grammatical structure and semantic information. GcGAT is a deep learning model, which introduces SAGPool, MLP, and Softmax based on GAT to classify vulnerable functions and non-vulnerable functions. Finally, the experiment results validate the correctness and effectiveness of our solution. However, our current work is limited to detecting vulnerable functions caused by function internal statements and system calls. In the future, we will study how to detect vulnerable functions under the existence of custom function calls that are more general in actual situations.

\section*{Acknowledgments}
This work was supported in part by the Key Research and Development Project of Sichuan Province (no. 2021YFG0160), the National Key Research and Development Program of China (no. 2019QY1406). The authors would like to thank the anonymous reviewers for their valuable comments and suggestions.

\end{document}